# Role of interference and entanglement in quantum neural processing

A.A.Ezhov

Troitsk Institute for Innovation and Fusion Research,
142092 Troitsk, Moscow Region, Russia
ezhov@triniti.ru

**Abstract**

The role of interference and entanglement in quantum neural processing is discussed. It is argued that on contrast to the quantum computing the problem of the use of exponential resources as the payment for the absense of entanglement does not exist for quantum neural processing. This is because of corresponding systems, as any modern classical artificial neural systems, do not realize functions precisely, but approximate them by training on small sets of examples. It can permit to implement quantum neural systems optically, because in this case there is no need in exponential resources of optical devices (beam-splitters etc.). On the other hand, the role of entanglement in quantum neural processing is still very important, because it actually associates qubit states: this is necessary feature of quantum neural memory models.

## Introduction

Quantum neural processing is a relatively new field based on principle idea both of quantum computation [1] and also of neural networks [2]. There have been proposed different approaches to the development of quantum neural networks, quantum associative memory and other neural-like systems utilizing the main quantum phenomena, such as quantum parallelism, states interference and entanglement [3]. It seems that, on contrast to the elaboration of quantum analogs of classical computing, development of quantum neural models should be performed taking into account principle differences between classical and neural computing. It can be suggested, that these differences permit to quantum neural computing to eliminate or, at least, to reduce some difficulties inherent to application of quantum idea to the classical computing. In particular, current discussion about the role of interference and entanglement for quantum computations should be re-examined and adjusted for quantum neural computing.

## Definition of quantum algorithm

Model of the quantum neural processing can be formulated using the definition of *quantum algorithm* given by Narayanan [4]. It suggests the implementation of following steps:

1. the problem is formulated in numerical form;
2. the initial configuration (quantum state) is determined;
3. the terminating condition is defined;

4. the problem is divided into smaller subproblems decided in parallel in different universes;
5. there must be some form of interaction (interference) between all universes, which yields the solution.

Apart from *interference* other central concept used in this definition is the concept of *universe*. So, many universes interpretation of quantum mechanics (MUI) is also used in this definition. It means that not only role of interference but also famous discussion on conceptual foundations of quantum mechanics has direct relation to given definition of the quantum algorithm. Despite of the necessity to touch briefly this discussion, main goal of current presentation is to estimate the role of interference in quantum neural processing.

**Interference and entanglement in quantum computing**

While the definition given above uses only the concept of *interference,* the phenomenon of *entanglement* is considered by remarkable group of the authors as crucial element of real quantum computation, which guarantees its preference to the classical computing [5,6]. Entangled states have unique properties, which, in particular, permit to create a noiseless channel for the transmission of quantum state in quantum teleportation [7], and de-localize quantum information in a system consisting of several subsystems (it permits to protect the system quantum state against destructive influence of environment using error-correcting codes [8,9]).

Interference of different alternatives (quantum parallelism) alone is considered by this group as insufficient, because if this superposition is factorizable, it also presents in conventional e.g. "optical" computing. It is also claimed, that analogous "non-quantum" computations can be performed using any classical waves. Note, that only feature of entangled state is that it is more or less *non-factorizable* superposition state. To determine general entangled state,

$$|\psi\rangle = \sum_{x=0}^{2^d-1} a_x |x\rangle,$$

it is necessary to define exponential number ($2^d$) of *independent* complex-valued amplitudes $a_x$[1]. Such general-form states are contrasted to the factorizing superposition

$$|\psi\rangle = (\alpha_0 |0\rangle + \beta_0 |1\rangle) \otimes ... \otimes (\alpha_{d-1} |0\rangle + \beta_{d-1} |1\rangle),$$

which can be described using only *2d* complex-valued amplitudes. Representing a non-entangled factorizing state in a conventional form $|\psi\rangle = \sum_{x=0}^{2^d-1} a_x |x\rangle$, it can be concluded that in this case the amplitudes $a_x$ are strongly correlated.

Because of entangled state is described by exponential number of parameters, calculation of function-dependent unitary transformation

---

[1] Really, this number is slightly less, because the state vector should be normalized. This normalization condition is a common amplitudes link both for entangled and non-entangled states.

$$|\psi\rangle|0\rangle \xrightarrow{U_f} \sum_{x=0}^{2^d-1} a_x |x\rangle|f(x)\rangle$$

can be done for all possible argument values *x* in parallel, giving the exponential speedup of calculations [5], if the function $x \mapsto f(x)$ can be computed *in polynomial time*. Obviously this is true for *any* superposition (this is the essence of quantum parallelism only), but *in common, this transformation makes the final state entangled*. It leads to the necessity to process exponential number of data while trying to perform analogous calculations classically. This will inevitably and exponentially decrease the speed of classical computations comparing with quantum computations [6]. But if all quantum states involved in computational process will be factorized (non-entangled), then they can be described by data, which amount grows only linearly with the number of qubits. In this case no speedup of quantum computing seems to be achieved. Corresponding calculations can be performed effectively using classical means. Taking into account such kind of considerations, just exponential speedup of calculations is considered by this group of authors as the main feature of *really quantum algorithms*. This speedup is strongly attributed to the entanglement of quantum states involved.

Using this consideration it is also argued, that any algorithms which do not use entangled states cannot give exponential decrease of any of computational resources – these resources are divided on *temporal*, e.g. the number of queries in Grover algorithm, *physical*, e.g. energy, measurement precision, number of physical devices used for algorithm implementation etc., and *readout*, e.g. number of quantum system copies needed to realize computation with given accuracy [10]. In this case it is claimed, that these algorithms or their implementations should not be considered as "really quantum".

As the result, a label "quantum" is sometimes removed not only from the implementations of quantum algorithms (including quantum state teleportation) based on the use of linear optics [11], because they use exponential number of physical devices (polarizers and beam-splitters), but also from all realizations of simplest variant of quantum algorithms of Deutsch-Jozsa [12-14] and Grover [15,16] using the NMR technique. This is because the existence of arguments in favor to the absence of the entanglement of qubit states imitated by states of different atoms of the molecules used in NMR experiments [17-19]. In the same manner quantum character of implementations based on the manipulations with Rydberg states of atoms [20,21] is also rejected, because they demand, besides others, the exponential accuracy of atom state measurements.

Hence, according to the described point of view to the essence of quantum computing, it is necessary either to use entangled states to perform real quantum computing, or, in the opposite case, to use exponential number of resources to compensate their absence. But in the last case corresponding computations cannot be called really quantum.

It seems however, that this position leads to the situation when the boundary between real quantum and pseudo quantum computing becomes fuzzy.

For example, it is known, that Deutsch algorithm, which really exponentially accelerates differentiation between balanced and constant Boolean functions [22], does not use entanglement in the case of the functions of one or two arguments [23] and can be implemented in these cases using linear optics. On the other hand, it really uses entangled states in the case of the functions of more variables [24]. It means, that Deutsch algorithm is in some sense "semi-quantum". It has been argued that Grover algorithm also does not demand entanglement in the case of d<3 qubits [14,25].

It can be easily recognized that Grover algorithm really demands entanglement of qubit states belonging to the key and informational registers. But it only establishes necessary associations between the states of these registers. Moreover, it actually demands the use of exponential resources for preparing the database quantum state, which inevitably collapses after any presented query set. Because of no-cloning theorem [26,27] there is no possibility to make efficiently many copies of this database. It will be necessary to use exponential number of operations to create every of them. It seems that maybe only Shor's algorithm really fits all demands of "strong definition" of quantum computing, because it definitely operates with entangled states and uses relatively simple function, which can be computed in polynomial time.

Other group of authors presents a number of examples of implementation of quantum algorithms, which do not use entanglement. One of the approaches proposed in [25] is based on the change the system consisting of many qubits with entangled states by the equivalent *single* quantum system, having exponential number of states. Because of the concept of entanglement has no sense for single system this approach turns to be universal for *disentanglement* of any quantum algorithm. However, as it was shown by D. Meyer [21], it again demands to perform exponentially precise measurement of quantum system states, so does not lead to really quantum computing. Note, that if the system considered has equally-spaced energy levels, then, instead of exponential precision, just the state energy will grow exponentially.

At the same time, D.Meyer also argued that the failure of Rydberg state manipulation *does not mean at all that real quantum computing cannot be performed using only interference of states* [28][2]. As example he described the algorithm initially proposed by Bernstein and Vazirani [30] and rediscovered by Terhal and Smolin [31]. This algorithm can be used for the search of given item in sophisticated database using single query.

However, as it can be easily seen, formulation of the search problem used in [28] is practically useless. It actually realizes some sort of "quantum sea battle', where single one-cell enemy ship have to be detected. On contrast, real search surely demands the entanglement of the key and informational registers (so we have to receive some information about searching ship) or the preparing a database (not a query) as a quantum state. Nevertheless, the use of Bernstein-Vazirani algorithms is convenient to illustrate the problem with the quantum implementation of Boolean functions. It is important to consider this implementation because it has direct correspondence to the efficiency of quantum approach applying to the solving of oldest neurocomputing problem.

Indeed, consider the quantum circuit realizing arbitrary Boolean function (Fig. 1). The initial database state

$$\frac{1}{2^{(d+1)/2}} \sum_{x=0}^{2^d-1} |x\rangle (|0\rangle - |1\rangle) |B(x)\rangle$$

---

[2] Recently Meyer's suggestion about the possibility to perform quantum computing using interference only was supported by experimental realisation of database search. Corresponding approach was called *wave computing* [29].

is prepared using the algorithm of Ventura and Martinez [32], which entangles all possible argument values $|x\rangle$ with the correspondent values of given Boolean function $B(x)$. Then, the transformation performed by Oracle – in this case the Oracle plays the role of *User* or *Similus Generator* (just it knows the presented argument value *a*) – is performed using a sequence of CNOT gates controlled only by qubits for which $a_i = 1$. For each $|x\rangle$ they will invert the sign of it as $|x\rangle \mapsto (-1)^{x \cdot a \bmod 2} |x\rangle \equiv (-1)^{x \cdot a} |x\rangle$ (note, that this system of CNOT gates simply counts the number of ones in $|x\rangle$ in those places where $a_i = 1$). It completes the second step of Bernstein-Vazirani algorithm. At last, the set of *d* Hadamard gates converts the quantum state to the final state $|\psi\rangle = |a\rangle \frac{1}{\sqrt{2}}(|0\rangle - |1\rangle)B(a)$ with the searched value of Boolean function in the last qubit.

It is evident, that entanglement turns to be necessary for the association of the argument and function values. However, although the entanglement of each argument-function pair of values can be done in polynomial time, using Ventura-Martinez algorithm, realization of arbitrary complex Boolean function will obviously demand exponential number of such steps $(2^d)$.

Obviously, one-query search of the Boolean function value cannot compensate the expenses of preparing the initial entangled state in the case of general form of function. So, temporal economy (in number of queries) is de-valued by the time needed for database entangled state preparing.

Another, more convenient approach, which uses the set of unitary transformation for realization of Boolean function will demand in the case of general form of function the use of exponential number of pulses – two qubit state exchanges (or exponential number of beam-splitters: any unitary 2x2 transformation using beam-splitters with input and output phase shifts [33,34], and any *N*x*N* transformation can be implemented using a matrix of these devices [35].

**Relation to quantum neural processing**

Such situation is not a problem of only quantum computing. Classical neural network realization of arbitrary Boolean function also demands the use of exponential number of hidden neurons and their interconnections. This exponential increase of resources, which is analogous to the exponential increase of the number of situations should be taken into account in solving problem using algorithmic approach, really turns neurotechnology to reject the goal of precise function realization and faces it with the problem of function approximation using not programming but learning on the restricted set of examples. The important point here is that the number of examples in training set should not grow exponentially with the problem size. Also, modern neural technology deals with analog signals, not with binary ones.

Taking into account the arguments presented above we can conclude that:

- Quantum analogs of classical neural systems have not to operate with the qubits, in such a manner as modern neural systems do not operate with bits, but process analog

- signals. It makes the problem of entanglement irrelevant for analog quantum neural systems.
- Quantum computing deals with the problem decision, not with function calculations as in neural computing. Even the realization of Boolean functions demands in common the use of exponential number of operations.
- The problem of exponential increase of resources with the size of input is not crucial for neural technology (both classical and quantum), because it does not use algorithmic approach connected with precise realization of functions, but approximate them, by training *modest* neural architecture on the *restricted number* of examples. Function approximation (really fitting restricted data points) is not computationally hard problem. Neural technology ordinary deals with *patterns* (wide-band signals) for which *d*>100, for which both Rent rule and also exponential grow of logical operations forbid the way of the use of exponential resources. Really, implementation of general Boolean function demands the use of exponential number of hidden neurons, $2^d$. But while training a network using *P* examples, this number grows as $\sqrt{P}$ for optimal 2-layer architecture *independent on the pattern's dimensionality*.
- Original difficulty of the modeling of quantum phenomena using classical computers is not inherently connected with the phenomenon of entanglement. It is due to the exponential number of situation, which should be taken into account in spite of Feynman path integral interpretation of quantum mechanics.
- The phenomenon of entanglement is nevertheless extremely important for those quantum neural system, which operate with qubits because it give the mean for association of the different part of quantum systems.

It also should be taken into account, while discussing possible physical implementations of analog quantum neural systems, that there is no *optical calculations per se*. There are exist only *classical* and *quantum* computations and different forms of their implementation, including optical. Note, that what is called optical computations do not used *light interference*. For example, optical implementations of classical neural networks use incoherent light of different wavelengths to implement interconnections between optical-electronic devices corresponding to model neurons. Recent developments in optical implementations of quantum algorithms are based on recognition of this fact [29].

General conclusion, which can be done, is that for the development of the model of analog quantum neural system [36] it is possible to use the definition of quantum algorithm and quantum computation given by Narayanan despite it does not use the concept of entanglement.

**On the role of many-worlds interpretations**

Other part of the definition of quantum algorithm given by Narayanan uses the concept of MUI of quantum mechanics. There exists an extensive discussion (pro and contra) of MUI [37-44]. Initially, arguments supporting MUI have been presented by specialists in quantum cosmology and later, in quantum computing. In *quantum computing* the advantages of this approach was successfully used and advocated by D. Deutsch [45,46]. In *quantum neural computing* it was used by A.Narayanan and T.Menneer [47]. It is not a goal of this presentation to discuss the meaning and value of MUI, because it is related to the principal and still unsolved problems of conceptual foundations of quantum mechanics. But it seems that there exists some reason to use MUI not only because of its evident convenience in applying to quantum computing and heuristic power[3] in development of different quantum-like algorithms (e.g. quantum Turing machine [46] or quantum genetic algorithm [48]), and not only because it can clarify easily the essence of analog quantum neural systems. It is possible to use it also, because it gives another look to the problem of resources, which can give the preference to quantum computations. It is also important to note, that Deutsch's variant of MUI is not operational but explanational theory. Just the *explanation* seems to be necessary for understanding the essence of quantum neural computing. Other important reason is that just *interference* (not entanglement) plays the central role in Deutsch's approach and Narayanan's definition of quantum algorithm. As it has been already stressed in previous section, just interference can be main feature of quantum neural processing.

First of all, in spite of the discussion of the role of entanglement, it is reasonable to note that Everett's original concept of *relative state* has direct correspondence to the concept of entanglement. In the same mode as it is impossible to assign definite quantum state to the separate qubit which is entangled with other qubit, it is impossible (according to Everett theory) to assign a well-defined state to the quantum subsystem interacting with the other subsystem. This state can be defined only relatively to the state of complementary subsystem, with which given one interacts. In interpretations of measurement process, just the outcome of the experiment is considered as related to corresponding state of the observer. So, the entanglement of the state of observer and device pointer, gives us other example of quantum-mechanical association.

In Everett theory only wave function evolved according Schrödinger equation can describe the universe evolution, no collapse phenomena happen, but after each "measurement-like interaction" our universe (really *multiverse*), is splitted on multiple copies in which *every* possible result of measurement (or interaction) is realized, and this resulting (basic) state is entangled with the corresponding state of observer (if he is present), or with the state of the interacting system.

Since Everett's pioneering work many variants of MUI have been developed. Some of them reject the existence of multiple real universes, considering only one as real, and other ones as subjective. Deutsch uses the variant of MUI, which forms single *multiverse*, representing the whole reality. According to this point of view all universes exist in parallel and can interfere with each other. This existence is considered as a root of quantum parallelism. In contrast to Everett, Deutsch suggests, that universes *do not split* but *differentiate* in measurement. So infinite number of parallel universes exists from the beginning. It should be noted, that many paradoxes of *quantum measurement* and also the essense of *counterfactual computations* can be easily explained just using MUI [49,50]. The main feature of MUI, which seems to be most important for quantum computing, is the possibility for the multiple universes *to interfere* with each other in a short time. Just the result of universes interference can be observed, making the quantum computing possible. Because our multiverse has enormous number of these copies there

---

[3] Opposite opinion can be also found: " … quantum computers are not wedded to "many worlds" interpretations, not only in terms of the prediction of the results of experiments, but also in terms of insight into what is going on within the quantum computational process" (Steane, 2000)

is a real *resource* for organizing such kind of interference (roughly speaking we can have enough copies of universe with enough number of *shadow* photons to organize the interference of arbitrary number of them). The only problem is how to use this resource (the possibility to utilize the particles in many universes) in order to provoke to interfere a lot of them.

On the other hand, it can be easily seen that modeling of many barrier and multi-slit interference does not demand exponential number of computational resources despite on the really exponential number of paths the photon can take (Fig. 2).

The amplitude of the probability of the photon to reach detector can be calculated classically in a short procedure. This procedure easily follows after the *imbedding* of the problem of calculation of amplitude for the photon to reach detector into the *family* of problems for it to reach each slit [51] – see Fig. 3. Some kind of invariance principle has also be used [52]. Here it follows from the absence of reflections (by suggesting), which permits to multiply amplitudes consequently. (The same approach is also used in *dynamic programming*)

It is clear, that this system can be described using $Nb$ parameters (where $N$ corresponds to the number of slits in each of $b$ barrier). It is just the situation of factorizable non-entangles quantum state.

*But it is not reasonable to reject this interference system, because in the case of neural networks applications it is not necessary to use exponential number of resources.*

In some sense, many-universes approach tells that even in the situation just considered exponential number of resources will factually be used. But it is not a problem, because they do not belong to "our universe".

Really, it can be said in MUI, that *exponential number* of photons from different universes will really interfere in the detector to produce the resulting picture. It seems rather non-practical to waste such amount of resources to perform so simple calculation, which should be performed on classical computer in a short time.

Really, each of these photons should calculate corresponding input to the whole amplitude not using any information about the considerable part of its computations performed by the *other* photon in the other universe (because "communications between different branches is impossible" [44]). This approach has been used in developing the model of quantum analog neuron, able to approximate any continuous function of many variables [36]. The ability of single neuron to calculate its output in many interfering "universes" permits eliminate the necessity to build *neural networks* to perform the same approximation problem [36]. This is the direct sequence of resource of interference (not of entanglement) which can be explained by MUI.

**Why quantum neural systems?**

But if there is no need in exponential resources in classical neural technology, why there is a need in quantum neural systems at all? In other words, what is the preferences of quantum neural systems to the classical ones?

There are many arguments in favor to quantum neural processing. Among of them are:
- the absence of pattern (classical) interference in content-addressable memories [48];

- the possibility to build exponential memory with exponential capacity [54]
- the possibility to reach better generalization by training the networks of simpler architectures;
- the possibility to eliminate quantum neurons in network ar all [35].
- etc.

It seems, that other very promising field of quantum neural processing is the development of trainable and adjustable quantum gates, which can be used in quantum computing, quantum control and quantum measurement suggesting the use of closed feedback loops. In these applications specially tuned laser pulses can be an outputs of quantum neuron system, which has as input a set of measurements of the state of quantum system which should be operated of this pulse. In this case the term "quantum" means that quantum neuron system really produces quantum system operating as gate for other quantum system. This clearly corresponds to the original function of artificial neural networks, which in principle could be used in $40^{th}$ as trainable logical gates for classical computers.

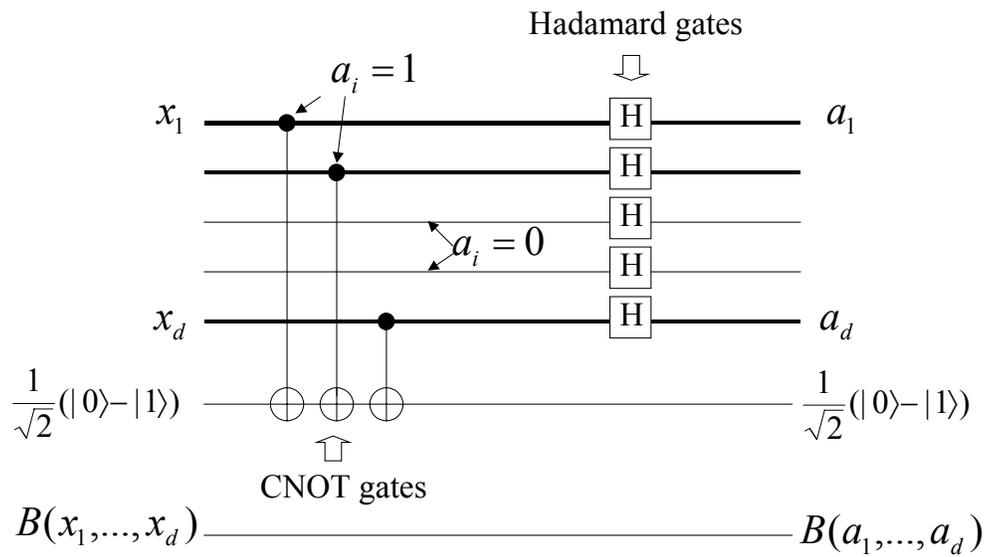

**Figure 1**. Quantum circuit performing calculation of Boolean function using Ventura-Martinez (VM) and Bernstein-Vazirani (BV) algorithms. The database quantum $d+2$ qubit state is the superposition of every d–qubit arguments $(x_1,...,x_d)$ entangled using VM algorithm with corresponding Boolean function value $B(x_1,...,x_d)$ represented by the state of $(d+2)$-th qubit. BV algorithm for single-query search operates with the first $d+1$ qubits, using a set of CNOT gates to implement dot product of external stimulus $a$ and argument value $x$. Final state represents the presented stimulus in the first $d$ qubits, and corresponding (entangled) value of Boolean function in $(d+2)$-th qubit

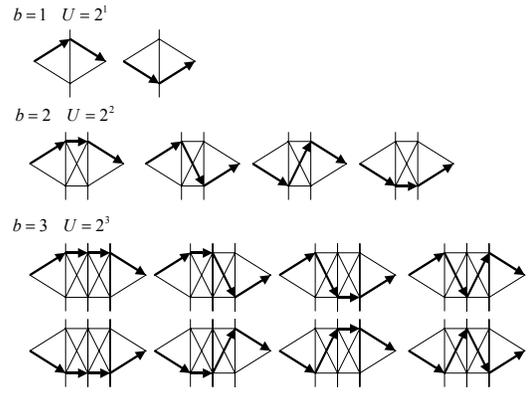

**Figure 2**. In multi barrier system (with 2 slits in each, in this example) the number of photon paths (universes) grows exponentially with the number of barriers, $b$.

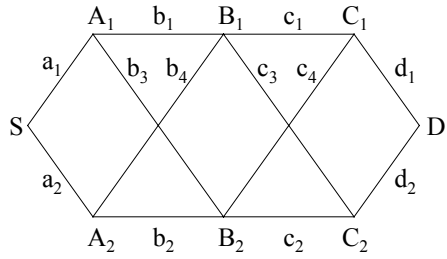

$$\langle D | S \rangle = \langle C_1 | S \rangle d_1 + \langle C_2 | S \rangle d_2$$

$$\langle C_1 | S \rangle = \langle B_1 | S \rangle c_1 + \langle B_2 | S \rangle c_4$$

$$\langle C_2 | S \rangle = \langle B_1 | S \rangle c_3 + \langle B_2 | S \rangle c_2$$

$$\langle B_1 | S \rangle = \langle A_1 | S \rangle b_1 + \langle A_2 | S \rangle b_4$$

$$\langle B_2 | S \rangle = \langle A_1 | S \rangle b_3 + \langle A_2 | S \rangle b_2$$

$$\langle A_1 | S \rangle = a_1$$

$$\langle A_2 | S \rangle = a_2$$

**Figure 3**. The initial problem of calculation of the amplitude to photon to reach detector $\langle D | S \rangle$ is imbedded into the family of problem for it to reach each slit $\langle L | S \rangle$, where $L \in \{A_{1,2}, B_{1,2}, C_{1,2}, D\}$. The amplitudes of each transition along a particular legs are defined by small characters. Then, the Cauchi problem for the definition of all of these amplitude can be formulated (right), which demand to use a number of operations which grows only linearly with the number of barriers.